\documentstyle{article}
\begin{document}
\LARGE
\vspace*{0.25in}
\begin{center}
\bf The Topological Origin of Black Hole Entropy 

\vspace*{0.6in}
\normalsize \large \rm 

Dept. of Physics

Beijing Normal University

Beijing 100875, China

Zhong Chao Wu

(Gravity Essay)

\vspace*{0.4in}
\large
\bf
Abstract
\end{center}
\vspace*{.1in}
\rm
\normalsize
\vspace*{0.1in}

In gravitational thermodynamics, the origin of a black hole's entropy is the topology of its instanton or constrained instanton. We prove that the entropy of an arbitrary nonrotating black hole is one 
quarter the sum of the products of the Euler characteristics of its horizons
with their respective areas. The Gauss-Bonnet-like form of the action is not only crucial for the evaluation, but also for the existence of the entropy. This result covers all previous results on the entropy of a nonrotating black hole with a regular instanton. The argument can be readily extended into the lower or higher dimensional model. The problem of quantum creation of such a black hole is completely resolved.

\vspace*{0.3in}

PACS number(s): 98.80.Hw, 98.80.Bp, 04.60.Kz, 04.70.Dy

\vspace*{0.3in}

e-mail: wu@axp3g9.icra.it

\pagebreak

Hawking radiation was the most important discovery in gravitational physics in the second half of the last century. From this scenario it is derived that 
the entropy of a Schwarzschild black hole is a quarter of its horizon 
area [1]. The entropy is interpreted as the measure of our ignorance about the
information beyond the horizon. In Euclidean quantum gravity, it was shown 
that the origin of the entropy is due to the nontrivial topology of its
Euclidean spacetime section, i.e., the instanton [2].

For many cases the concrete relation of the entropy and topology of a black hole has been worked out [3][4]. It is widely believed that the entropy of a nonrotating black hole is one 
quarter the sum of the products of the Euler characteristics of its horizons involved
and  their respective areas. However, this has only been proven for the case of a black hole with a regular instanton. For all known nonrotating black hole cases, a regular instanton can be obtained in three cases: (i) when only one horizon is involved, or (ii) two horizons with the same surface gravities $\kappa$ are involved, or (iii) at least one of the two horizons involved recedes into an internal infinity. Only very recently, this relation for an arbitrary nonrotating black hole without a regular instanton has been proven [5]. 
This allows black holes with regular instantons to be a special case of our considerations here.

In gravitational thermodynamics [6][7], it is known that for a generic black hole, one can still define a temperature as $\kappa/2\pi$ associated with each horizon. However, there does not exist a thermal equilibrium state with an uniform temperature. Or equivalently, there does not exist a regular instanton. This implies that neither
canonical nor grandcanonical ensemble can apply here. Fortunately, one can
circumvent this obstacle using microcanonical ensemble. In
contrast, for microcanonical ensemble, the temperature is not
defined, but all conserved quantities are given. Since the
probability of each state
under the conserved quantity constraints are equal, the entropy
is simply the
logarithm of the number of these states. 
 
The partition function for microcanonical  ensemble in gravitational thermodynamics is the
path integral
\begin{equation}
Z = \int d[g_{\mu \nu}]d [\phi]\exp-I
\end{equation}
where the path integral is over all spacetime metrics $g_{\mu
\nu}$ and matter configurations $\phi$ under the conditions for the `equator'
imposed by the restriction of the microcanonical ensemble, and $I$ is the Euclidean action.

The Euclidean action in the Einstein theory is [2]
\begin{equation}
I = - \frac{1}{16 \pi} \int_M \sqrt{g}(R - 2 \Lambda + L_m) 
d^4x- \frac{1}{8\pi}
\oint_{\partial M }\sqrt{g} Kd^3 x,
\end{equation}
where $R$ is the scalar curvature of the spacetime manifold $M$,
$K$ is the expansion rate of the boundary $\partial M$,
$g$ is the determinant of the metric for the 4-metric or its
lower dimensional version, $\Lambda$ is the cosmological
constant, and $L_m$ is the Lagrangian of the matter content.

The main contribution to the path integral is
from a stationary action orbit. The partition function is
approximated by the exponential of the negative of the action of
the orbit. This is called the $WKB$ approximation, which we shall use
in this paper. 

If the action of the orbit is stationary with respect to all
variations, then one obtains an instanton. It is determined
essentially by the topological properties of the manifold. The
metric is regular without any singularity. 

On the other hand, the dominating contribution to the partition function for the microcanonical ensemble is called constrained instanton [8][9]. It is a manifold for which the action is stationary with respect only to the variations under the restrictions due to the ensemble, instead of with respect to all variations under no restriction. 

For the quantum creation scenario in the no-boundary universe [10], the relative probability  takes the same form (1) and the path integral is over all 4-metrics with the  given 3-metric and matter content at the equator. These constraints can be
characterized by a few parameters, like mass $m$, charge $Q$ and
angular momentum $J$ for the black hole case. These conditions
are exactly the same as for
the microcanonical ensemble in gravitational thermodynamics. Therefore, the constrained instanton is also the creation seed in the no-boundary universe. The exponential of the negative of the instanton action is the relative creation probability of the universe. 

Since the entropy is the logarithm of the partition function in
microcanonical ensemble, then at the
same level, the entropy is the negative of the action.

For all cases of black holes considered, the
spacetime has a $U(1)$ isometry. The group parameter is
identified as the Killing time coordinate. The Euclidean nonrotating black hole metric takes the form
\begin{equation}
ds^2 = \Delta (r) d\tau^2 + \Delta^{-1}(r) dr^2 + r^2
d\Omega^2_2,
\end{equation}
where $\tau = it$, and the 2-metric $d\Omega^2_2$ is a compact
manifold which
does not depend on coordinates $\tau$ and $r$. For an ordinary black hole 
the 2-metric is a sphere, and for a topological black hole it is a 
compactified plane or hyperboloid. The zeros of the
rational function $\Delta (r)$ are the horizons. 

One can construct a compact constrained instanton (to be justified below) using a sector
between two horizons  denoted by two zeros $r_l, r_k$ in the
identified manifold. The surface gravity $\kappa_i$ of the
horizon $r_i$ is $- d \Delta (r)/2dr| _{r = r_i}$. If the zero is
of multiplicity 1, then one obtains a nonzero $\kappa_i$. On the
two dimensional space $(\tau, r)$ the conical singularity at the
horizon can be regularized by choosing $\beta = 2\pi \kappa_i
^{-1}$.  For two horizons with same nonzero surface gravities,
one can use the same $\beta$ to obtain a compact regular
instanton. If these two surface gravities are different, then the
constrained instanton has at
least one conical singularity at the horizons, since no value of
parameter $\beta$ can regularize both of the horizons
simultaneously. The de Sitter model is an exception, since $r$ is
identified with $-r$, only one horizon, i.e. the cosmological
horizon is needed for the construction of the instanton.

If one of the two zeros is of multiplicity larger than 1,
then its surface gravity $\kappa_i = 0$, and the associated
horizon recedes to an internal infinity. Then it is always
possible to regularize the other
horizon by choosing a right value $\beta$ to obtain a regular
instanton. The most familiar case is the extreme
Reissner-Nordstr$\rm\bf\ddot{o}$m black hole in the nonvacuum
model [4].

Now, let us calculate the action of the pasted manifold. We
use $M_l$ to denote the small neighbourhood of horizon $r_l$ with
the boundary of a constant coordinate $r$. The Euler number
$\chi(l)$ for the 2-dimensional $(\tau, r)$ section of
neighbourhood with zero (nonzero) surface gravity is 0 (1). We
use $M^\prime$ to denote $M$ minus $M_l$ and $M_k$. For the form
of action (2), the total action is the sum of those from the
three submanifolds.

First of all, let us consider the vacuum model with 
a cosmological constant. The total action is [11][12]
\begin{equation}
I = I_l + I_k + \int_{M^\prime} (\pi^{ij}\dot{h}_{ij} - NH_0 -N_i
H^i)d^3x d\tau,
\end{equation}
where the actions $I_l$ and $I_k$ are the actions for $M_l$ and
$M_k$. The action of $M^\prime$ has been recast into the
canonical form. $N$ and $N_i$ are
the lapse function and shift vector, $h_{ij}$ and $\pi^{ij}$ are
the 3-metric and the conjugate momenta respectively, $H_0$ and
$H^i$
are the
Einstein and momentum constraints, and the dot denotes the time
derivative. The manifold satisfies the Einstein
equation, and all time derivatives vanish due to the $U(1)$
isometry, therefore the integral over $M^\prime$
is equal zero.

Now the action $I_l$ or $I_k$ can be written
\begin{equation}
I_i =- \frac{1}{16 \pi} \int_{M_i}\sqrt{g} (R - 2 \Lambda)d^4x
- \frac{1}{8\pi} \oint_{\partial M_i }\sqrt{g} K d^3x,\;\; (i =
l, k).
\end{equation}
If there is a conical singularity at the horizon, its contribution to the action
can be considered
as the degenerate version of the second term of the action, in
addition to that from the boundary of $M_i$. The conical
singularity 
contribution is termed as a deficit ``angle'' due to its
emergence. If the horizon recedes into an internal infinity, then
this is no longer of concern.

One can apply the Gauss-Bonnet theorem to the 2-dimensional
$(\tau, r)$ section of $M_i$,
\begin{equation}
 \frac{1}{4 \pi} \int_{\hat{M}_i}\sqrt{g}^2Rd^2x +
\frac{1}{2\pi}
\oint_{\partial \hat{M}_i }\sqrt{g} ^1K d^1x +
\frac{\delta_i}{2\pi} = \chi (i),
\end{equation}
where $\hat{M}_i$ is the projection of $M_i$ onto the 
2-dimensional $(\tau, r)$ section, $^2R$ is the scalar curvature
on
it, $^1K$ is the corresponding expansion rate, $\delta_i$ is the
total deficit angle, and
$\chi(i)$ is the Euler characteristic of $\hat{M}_i$. Since the
expansion rate of the
subspace $r^2 d\Omega^2_2$ goes
to zero at the horizon, $K$ and $^1K$ are equal. Comparing eqs
(5) and (6), one can see that as the
circumference of the boundary tends to infinitesimal, the action
(5) becomes $-\chi(i) A_i/4$, where $A_i$ is the surface area of
the horizon. It is noted that both the volume integral of (5) and
the first term of the left hand side of (6)  vanish as the
boundary approaches the horizon. The same result is obtained
regardless of the existence of the conical singularity at the
horizon or not, i.e., it is independent of the value $\beta$.
Here the conical singularity contribution is represented
by $\delta_i/2\pi$.

From (4)-(6), we learn that the action is independent of the
parameter $\beta$. Since the manifold satisfies the Einstein equation
everywhere with probable exception at the horizons. The conical singularities
there are parametrized by $\beta$, the only degree of freedom left.
Therefore, the manifold is qualified as a constrained instanton.
The entropy, or the negative of the total action of the
constrained instanton is
\begin{equation}
S = - I =  \frac{1}{4} ( \chi(l) A_l + \chi(k) A_k).
\end{equation}
This is a quite universal formula.

If one use $r_0$ to denote the maximum zero of $\Delta(r)$ and
the metric of the sector $r > r_0$ is Euclidean, then this sector
can also be used for constructing an open instanton. 

The action of the open instanton is divergent. One can modify the action form (4) to regularize it, by setting $I_l = I_0$, then dropping $M_k$ term and finally letting $M^\prime$ be the sector $M$ minus $M_0$. 
The form of this action is derived from the
requirement that, as the mass $m$ (and the angular momentum $J$
for a
rotating model not expressed by (3)) is
held fixed at infinity with an appropriate asymptotic falloff
for the field, the Einstein and field equations must be derived
from the action [13]. These boundary conditions correspond to the
microcanonical ensemble. Follow the same argument, one can
derive 
$S =\chi(0) A_0/4$. 

One can include the Maxwell or gauge field into the model. For the magnetic black hole, the Maxwell action is compatible with the requirement of the microcanonical ensemble. On the other hand, for the electric black hole, it is not. One has to appeal to a Legendre transformation here. In both cases, formula (7) remains valid [5].

It is noted that the action of the black hole metric must be a linear function of the parameter $\beta$ due to the $U(1)$ isometry. The independence of the action from the parameter is at the edge of the knife. Indeed, the Gauss-Bonnet-like form is crucial not only for the evaluation, but also for the existence of the entropy! The Equivalence Principle is not sufficient to restrict the action to take the form (2) in the 4-dimensional spacetime. The deep reason behind this remains a mystery in Nature.

The method of the dimensional continuation of the Gauss-Bonnet
theorem has been used to study the  entropy of a black hole with
a regular instanton [3][14]. The formula presented in [15] does not apply to a topological black hole [16][17]. In
contrast, formula (7) is true for all nonrotating black holes.

Our analysis can be generalized into the Lovelock theory of
gravitation [18]. One can study $n-$dimensional black holes, which are described by (3) with $d\Omega^2_2$ replaced by $d\Omega^2_{n-2}$. 
The entropy of a black hole or the negative of the action is [5]
\begin{equation}
S = - I =  \frac{1}{4} ( \chi(l) A_lf_l + \chi(k) A_kf_k),
\end{equation}
where $f_i$ is a numerical factor determined by the  metric $r^2d\Omega^2_{n-2}$ of the horizon.

The discussion can be extended straightforward to the
nonvacuum  model. The  formula for the entropy of a black
hole with a regular instanton has been previously obtained
[20][21][14]. The constrained instanton can be used as a seed for quantum
creation of a black hole in Einstein gravity and its higher or even lower dimensional black hole [22]. However, one has to use the instanton with the largest action [9].  After we have obtained 
formulas (7) and (8), we no longer need to check stationary
property of the constrained
instanton case by case for an arbitrary black hole described by (3) and its lower or higher dimensional version.  The entropy and quantum creation of an arbitrary nonrotating black hole are thus completely resolved.

\vspace*{0.3in}
                       
\bf References:

\vspace*{0.1in}
\rm

1. S.W. Hawking, \it Commun. Math. Phys. \rm \underline{43}, 
199 (1975).

2. S.W. Hawking, in \it General Relativity: An Einstein
Centenary Survey, \rm eds. S.W. Hawking and W. Israel, (Cambridge
University Press, 1979).

3. G.W. Gibbons and R.E. Kallosh, \it Phys. Rev. \rm \bf D\rm
\underline{51}, 2839 (1995).

4. S.W. Hawking, G.T. Horowitz and S.F. Ross, \it Phys. Rev. \bf
D\rm\underline{51}, 4302 (1995).

5. Z.C. Wu, gr-qc/9911078, \rm (in press).

6. G.W. Gibbons and S.W. Hawking, \it Phys. Rev. \bf D\rm
\underline{15}, 2752 (1977).

7. G.W. Gibbons and S.W. Hawking, \it Phys. Rev. \bf D\rm
\underline{15}, 2738 (1977).

8. Z.C. Wu, \it Int. J. Mod. Phys. \rm \bf D\rm\underline{6}, 199
(1997), gr-qc/9801020.

9. Z.C. Wu, \it Phys. Lett. \rm \bf B\rm
\underline{445}, 274 (1999), gr-qc/9810012.

10. J.B. Hartle and S.W. Hawking, \it Phys. Rev. \rm \bf D\rm
\underline{28}, 2960 (1983).

11. C. Teitelboim, \it Phys. Rev. \rm \bf D\rm
\underline{51}, 4315 (1995).

12. S.W. Hawking and G.T. Horowitz, \it Class. Quant. Grav. \rm
\rm\underline{13}, 1487 (1996).

13. T. Regge and C. Teitelboim, \it Ann. Phys. \rm (N.Y.)
\underline{88}, 286 (1974).

14. M. Ba$\tilde{n}$\rm ados, C. Teitelboim and J. Zanelli, 
\it Phys. Rev. Lett. \bf \rm \underline{72}, 957 (1994).

15. S. Liberati and G. Pollifrone, \it Phys. Rev. \bf D\rm
\underline{56}, 6458 (1997). 

16. D.R. Brill, J. Louko and P. Peldan, \it Phys. Rev. \rm \bf
D\rm \underline{56}, 3600 (1997).

17. Z.C. Wu, gr-qc/9907064.

18. D. Lovelock, \it J. Math. Phys. \rm \underline{12}, 498
(1971).

19. C. Teitelboim and J. Zanelli, \it Class. Quant. Grav. \rm
\underline{4}, L125 (1987).

20. T. Jacobson and R. Myers,  \it Phys. Rev. Lett. \bf \rm
\underline{70}, 3684 (1993).

21. R.C. Myers and J.Z. Simon, \it Phys. Rev. \rm \bf D\rm
\underline{38}, 2434 (1988).

22. Z.C. Wu, gr-qc/9907065.

\end{document}